\documentclass{article}

\usepackage{arxiv}

\usepackage[utf8]{inputenc} % allow utf-8 input
\usepackage[T1]{fontenc}    % use 8-bit T1 fonts
\usepackage{hyperref}       % hyperlinks
\usepackage{url}            % simple URL typesetting
\usepackage{booktabs}       % professional-quality tables
\usepackage{amsfonts}       % blackboard math symbols
\usepackage{nicefrac}       % compact symbols for 1/2, etc.
\usepackage{graphicx}
\usepackage{amssymb}
\usepackage{amsmath}
\usepackage{lineno}
\usepackage{listings}
\usepackage{subfigure}
\usepackage{comment}
\usepackage{enumitem}
\usepackage[dvipsnames]{xcolor}

\title{Transfer learning for self-supervised, blind-spot seismic denoising}

\author{
  Claire Birnie \\
  KAUST\\
  Thuwal, Kingdom of Saudi Arabia \\
  \texttt{claire.birnie@kaust.edu.sa}
   \And
  Tariq Alkhalifah \\
  KAUST\\
  Thuwal, Kingdom of Saudi Arabia \\
    }
    
\begin{document}

\chead{Transfer learning for seismic denoising}

\maketitle

\begin{abstract}
Noise is ever present in seismic data and arises from numerous sources and is continually evolving, both spatially and temporally. The use of supervised deep learning procedures for denoising of seismic datasets often results in poor performance: this is due to the lack of noise-free field data to act as training targets and the large difference in characteristics between synthetic and field datasets. Self-supervised, blind-spot networks typically overcome these limitation by training directly on the raw, noisy data. However, such networks often rely on a random noise assumption, and their denoising capabilities quickly decrease in the presence of even minimally-correlated noise. Extending from blind-spots to blind-masks has been shown to efficiently suppress coherent noise along a specific direction, but it cannot adapt to the ever-changing properties of noise. To preempt the network's ability to predict the signal and reduce its opportunity to learn the noise properties, we propose an initial, supervised training of the network on a frugally-generated synthetic dataset prior to fine-tuning in a self-supervised manner on the field dataset of interest. Considering the change in peak signal-to-noise ratio, as well as the volume of noise reduced and signal leakage observed, using a semi-synthetic example we illustrate the clear benefit in initialising the self-supervised network with the weights from a supervised base-training. This is further supported by a test on a field dataset where the fine-tuned network strikes the best balance between signal preservation and noise reduction. Finally, the use of the unrealistic, frugally-generated synthetic dataset for the supervised base-training includes a number of benefits: minimal prior geological knowledge is required, substantially reduced computational cost for the dataset generation, and a reduced requirement of re-training the network should recording conditions change, to name a few. Such benefits result in a robust denoising procedure suited for long term, passive seismic monitoring.
\end{abstract}

\section{Introduction}
% Noise is bad, particularly for microseismic monitoring
Noise is a constant, yet undesirable, companion of our recorded seismic signals. Noise arises from a variety of sources, ranging from anthropogenic activities to industrial endeavours to undesirable waves excited by the seismic source (e.g., groundroll occurring during a reflection seismic survey); each of these sources generate noise signals with their own characteristics (energy and frequency content) and duration. Therefore, the composition of the overall noise field observed in seismic data, often just referred to as noise, is continuously changing. Noise is particularly troublesome for microseismic monitoring where the signal often arises from a low magnitude event, and is therefore, hidden below the noise \cite{Maxwell2014}, which can result in errors and artefacts in detection and imaging procedures \cite{Bardainne2009}. As such, large efforts are made to reduce the noise from data collection (e.g., \cite{Maxwell2010,Auger2013,Schilke2014}) to data processing (e.g., \cite{Eisner2008,Mousavi2016,Birnie2017}), and to ensure monitoring algorithms are adequately tested under realistic noise conditions \cite{Birnie2020}.

% DL challenges for noise suppression & rise of SSL
Following the great success in several scientific domains, in the last decade Deep Learning (DL) has seen a rising interest in the geophysics community. As far as seismic denoising applications are concerned, \cite{Saad2020} utilised an autoencoder for random noise suppression, \cite{Kaur2020} utilised cycleGANS for groundroll suppression, and \cite{Xu2022} used deep preconditioners for seismic deblending, among others. One major challenge of such DL procedures is that they are trained in a supervised manner and therefore require pairs of noisy-clean data samples for training - an often unobtainable requirement in seismology. Whilst some studies have investigated the use of synthetic datasets for network training, this introduces uncertainty when applying the network to field data due to the large difference between field and synthetic seismic data \cite{Alkhalifah2021}. In an attempt to reduce this difference, many exert a great deal of effort in generating `realistic' synthetic datasets, which often require costly waveform and noise modelling. Another alternative, some geophysicists have adopted, is to denoise the field data using conventional denoising methods, like  local time–frequency muting filters, to generate the clean target data (e.g., \cite{Kaur2020}). Whilst such an approach may speed up the final denoising procedure (due to fast inference of neural networks), the performance of the trained DL denoising procedure will, at best, equal that of the conventional method used for generating the training labels.

As the requirement of having direct access to noisy-clean data pairs is unfeasible across many scientific fields, \cite{Krull2019} proposed the use of self-supervised, blind-spot networks for random noise suppression; by training a network directly on single instances of noisy images, they demonstrated promising denoising performance on both natural and microscopy images. \cite{Birnie2021} adapted the methodology of \cite{Krull2019} for suppression of random noise in seismic data, providing examples of successful applications to both synthetic and field data. However, the study of \cite{Birnie2021} also highlighted the degradation in the denoising performance as any correlation begins to exist within the noise. Building on this, \cite{Liu2022,Liu2022a} proposed to extend the blind-spot property to become a blind-trace, adapting the previously proposed self-supervised denoiser for the suppression of trace-wise noise. Similarly, both \cite{Luiken2022} and \cite{Wang2022} proposed blind-trace networks implemented at the architecture level, as opposed to the likes of \cite{Krull2019, Birnie2021,Liu2022} which were implemented as processing steps. Both \cite{Liu2022} and \cite{Luiken2022} illustrated successful suppression of trace-wise noise, specifically poorly coupled receivers in common shot gathers and blending noise in common channel gathers, respectively. However as highlighted in numerous studies, noise is continually changing and is neither fully random nor fully structured and, therefore, cannot be represented via a single distribution \cite{Birnie2016}. Earlier implementations of self-supervised, blind-spot networks have all focussed on suppressing a single component of the noise field as opposed to tackling the noise field as a whole.

% Applications of TL in seismology
In an attempt to target the noise field as a whole, we propose the use of transfer learning to boost the performance of these self-supervised procedures. Transfer learning involves pre-training a network on one specific dataset, or task, and using the network weights to warm-start the training process on either a new dataset, or a new task \cite{Torrey2010}. Within geophysics, a number of studies have already highlighted the potential of transfer learning. For example, \cite{Wang2021a} proposed the use of transfer learning for a seismic deblending task where an initial network is trained on synthetic data and fine-tuned on labelled field data obtained via conventional deblending methods. Similarly, \cite{Zhou2021} pre-trained a model with synthetic data prior to fine-tuning on field data, however this time the task at hand was fault detection. Both these studies illustrated the benefit of pre-training on synthetic data despite being slightly constrained by the need for some labelled field data. Within this study, we propose supervised pre-training on a `frugally' generated synthetic dataset prior to using the trained weights for initialising a self-supervised network that we train on the field data.

% Conc
Self-supervised, blind-spot networks are one of the very few options that allow training and applying a network utilising only noisy field data. Given adequate time, such networks learn to predict all the coherent components that contribute to a central pixel's value, both signal and noise. Focussing on a microseismic use case, in this work we illustrate how blind-spot networks can be pre-trained on synthetic data to learn to predict the seismic signal from noisy data prior to fine-tuning the network for a reduced number of epochs to learn to reproduce the field seismic signature; stopping the network's training before it learns to replicate the coherent noise present in the field data. The utilisation of transfer-learning allows us to tackle the full noise field without pre-defining noise structures, as required in previous blind-spot procedures. In these experiments, we explicitly use frugally generated synthetic datasets for the network pre-training, removing the computationally expensive - and often non-trivial - task of generating realistic synthetic datasets. Following supervised training on the synthetic dataset, the network is further trained in a self-supervised manner to adapt to the seismic signal observed in field data. Benchmarked against both supervised and self-supervised procedures, the supervised synthetic base-training followed by the field self-supervised training is shown to improve the SNR of the data whilst introducing the least damage to the desired signal. Furthermore, the inclusion of the pre-training allows a tuning of the denoised solution ranging from a cleaner product with a smoothed signal to a high-quality signal with a little more noise, depending on our needs, which may be driven by down-the-line tasks of interest.

%%%%%%%%%%%%%%%%%%%%%%%%%%%%%
\section{Theory: Self-supervised, blind-spot networks}
There are two key terminologies repeatedly discussed in this paper:
\begin{itemize}
    \item \textbf{self-supervised learning}: where parts of the same training sample are used as both the input and target for training a neural network (NN), and
    \item \textbf{blind-spot networks}: a specific NN implementation where a central pixel's value is hidden in the input to the NN for training therefore requiring the network to learn most of the pixel's value from the neighbouring pixels.
\end{itemize}
Typically, blind-spot networks are implemented in a self-supervised manner. However, there is no technical reason as to why blind-spot networks cannot be implemented in a supervised manner, as will be done later in this paper. For the remainder of this section, we will introduce the theory of blind-spot networks from self-supervised point of view.

Conventional deep learning denoising methodologies rely on having pairs of clean and noisy samples for training. In some instances this may be possible, such as using photos on a perfectly clear day to train a network to suppress rain noise observed at the same location on a subsequent day \cite{Ren2019}. However, in many fields such as microscopy, medical imaging and geoscience, it is almost impossible to collect perfectly clean images to be used as the training target. For MRI, \cite{Lehtinen2018} recast this problem to overcome the requirement for a noisy training target. Assuming they had pairs of images with the same underlying signal but different noise properties, they showed how a network can be trained to map between the two images. However, as the noise varies between samples, the network never truly learns to map the noise and the resulting prediction is just the signal. Whilst removing the requirement of a clean target, this approach still required multiple instances with an identical signal - an unobtainable ask for wave-based data such as those obtained from seismic monitoring.

\begin{figure}[ht]
  \centering
  \includegraphics[width=.35\textwidth]{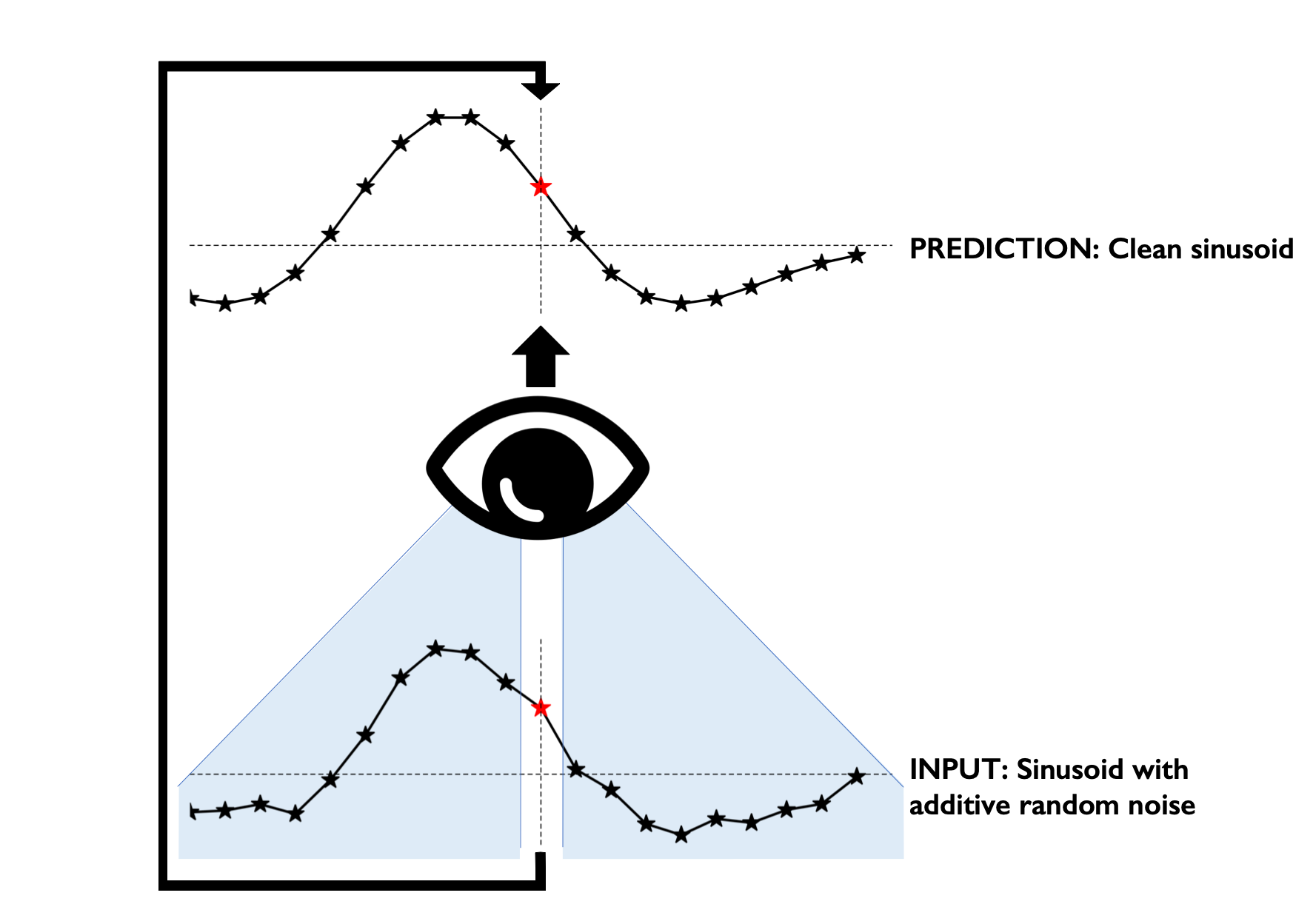}
  \caption{Schematic illustration of a noisy trace undergoing blind-spot denoising. For the prediction of a central value (red star), the input to the network is the nearby samples (blue area). The central value is not used in the prediction of its own value.}
  \label{fig:BStrace}
\end{figure}

Self-supervised, blind-spot networks were originally proposed as a means to overcome this requirement of having pairs of training data. \cite{Krull2019} identified that under the assumption that signal is coherent and noise is independent and identically distributed (i.i.d.), then the domain mapping of \cite{Lehtinen2018} could be extended to utilising a single instance per training sample. To do so, the network must use only neighbouring pixels to predict a central pixel's value and cannot be exposed to the original, noisy value of the central pixel. Figure \ref{fig:BStrace} illustrates what a blind-spot network would be exposed to when predicting a central value on a noisy trace. Assuming the trace is contaminated by random noise, the network cannot learn to predict the noise component of the trace and, therefore, only the signal component will be reproduced. 

\begin{figure}[ht]
  \centering
  \includegraphics[width=1.\textwidth]{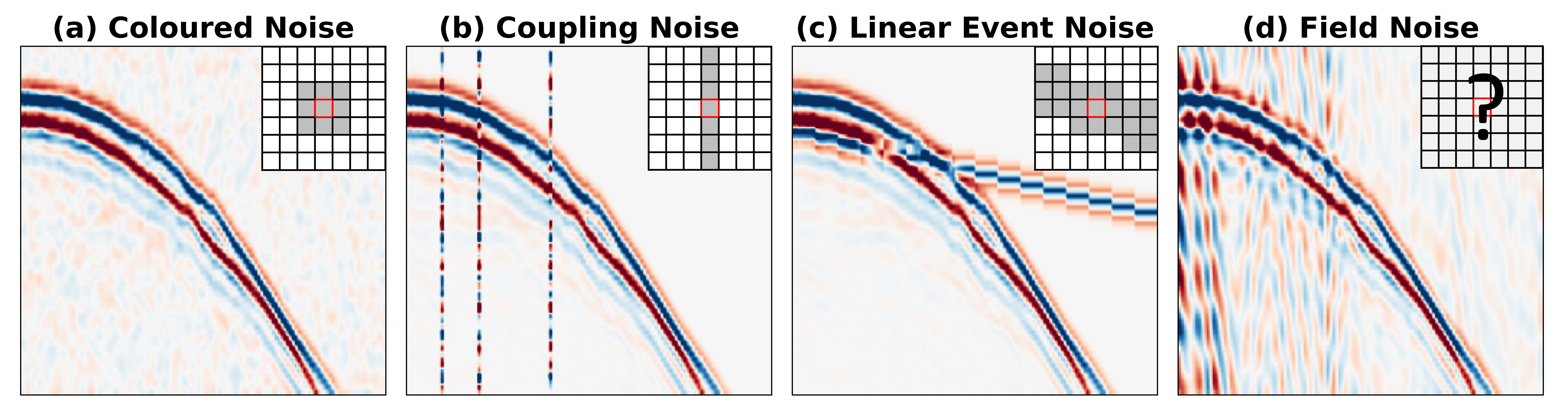}
  \caption{Different noise types and their respective blind-masks. (a) Coloured noise, (b) trace-wise noise, (c) noise from a far away event, and (d) field noise.}
  \label{fig:StructN2V}
\end{figure}

Extensions to blind-spot networks have recently been proposed allowing the suppression of coherent noise along a specific direction. \cite{Broaddus2020} originally proposed an extension of the `blind-spot', termed Structured Noise2Void, to mask the full area of coherent noise that may contribute to a pixel's predicted value. Recall - the network learns to predict the non-random components of the central pixel. Therefore, without the masking of coherent noise, the network would learn to replicate both the signal and noise. Figure \ref{fig:StructN2V} shows how the blinding structure could be constructed for different noise types observed in seismic data. The first (a) is noise with a short-lived correlation in time and space, possibly due to a local meteorological phenomenon. The mask for such noise, assuming equal spatio-temporal correlation, is just an extension of the blind-spot to a blind-square. The second (b) noise type shown is trace-wise, arising from a handful of poorly-coupled receivers. In this instance, the noise is coherent across the trace so the blind-spot is extended to a blind-trace, as proposed by \cite{Liu2022}. The next noise type (c) is the arrival of a seismic wave from a distant event, causing a linear move-out pattern. In this case, the blind-spot becomes a blind-line with the same rotation angle as the expected event. The challenge with building extensions of the blind-spot is that they require constant noise properties across the full training set. 

Whilst it is vital for successful denoising to remove the presence of coherent noise, it is important to note that by removing those pixels' contributions from the network's input can result in a substantial reduction in the amount of signal the network will see. Considering the final example (d) in Figure \ref{fig:StructN2V}, the properties of the coherent noise affecting one pixel to the left of the array are significantly higher to those affecting a pixel in the middle which are again significantly different to those on the left of the array. As such, it is highly non-trivial to design a blind-mask that would stop all coherent noise being provided to the network whilst still permitting enough signal information. For this reason, we return to the initial blind-spot approach and consider how we can adapt the training procedures to reduce the influence of coherent noise without explicitly masking it out.

In this work, we assume that it is impossible to obtain perfectly clean samples of field data for training and that there will always be a difference in features between synthetic and field data. Therefore, if we wish to train on data with the same properties as the data onto which we intend to apply the trained network then we must use the noisy, raw field data as our training input and target. As mentioned in the introduction, the `blinding' operation can be achieved through data manipulation (e.g., \cite{Krull2019}) or network design (e.g., \cite{Laine2019}). A comparison of the two approaches has not yet been published to identify if one or the other is better for seismic data. Therefore, following on from the study of \cite{Birnie2021}, we utilise the pre-processing method, which will be discussed in the following methodology section. 

%%%%%%%%%%%%%%%%%%%%%%%%%%%%%
\section{Methodology}
Self-supervised, blind-spot networks have been shown to be strong suppressors of i.i.d. noise without requiring pairs of noisy-clean training samples. However, the assumption under which these networks were proposed breaks as soon as any correlation exists within the noise \cite{Krull2019}. In this section, we first introduce our implementation of blind-spot networks through pre-processing alongside a tailored loss function. This is followed by a discussion on the general neural network (NN) architecture and training schemes used in this study and a detailed description of the implementation of transfer learning in this application.

Going forward, the term supervised training refers to the scenario where noisy data are used as input to the NN and clean data are the NN's target. Whilst self-supervised training refers to the scenario where no clean data are used, therefore a modified version of the noisy data are used as the input of the NN and the raw, noisy data are the NN's target. Blind-spot networks specifically refer to an adaptation of conventional networks where a randomly chosen pixel has its value removed from the network's receptive field, forcing the network to learn to predict this pixel's value from neighbouring pixels. Herein, the randomly chosen pixels will be referred to as active pixels, referring to their role in the loss function used for training the network. As highlighted below, blind-spot networks can be implemented in both a supervised and self-supervised fashion.

\subsection{Blind-spot implementation}
\begin{figure}[ht]
  \includegraphics[width=1.\textwidth]{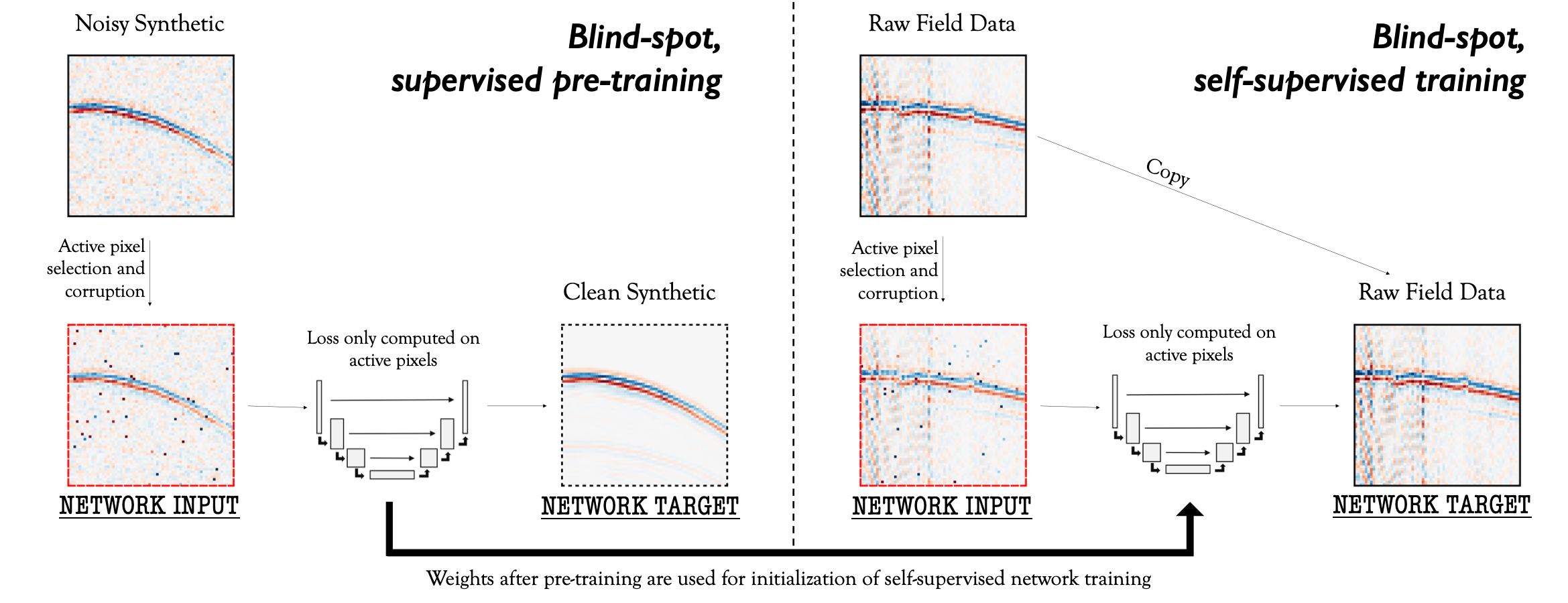}
  \caption{Proposed workflow for the incorporation of transfer learning into self-supervised, blind-spot networks. First, a blind-spot network is trained in a supervised fashion with a clean target (left). The weights after training are then used to initialise a new network, which is trained directly on the field data, in a self-supervised manner.}
  \label{fig:workflow}
\end{figure}

Under the assumption that signal is correlated between nearby pixels and that noise is i.i.d., a model can be trained to predict the signal component of an active pixel based off neighbouring pixels' values. To ensure the network cannot use the active pixel to predict itself, a pre-processing step is implemented that identifies a number of these active pixels and replaces their value with that of a neighbouring pixel, as illustrated in Figure \ref{fig:workflow}. In our implementation, this active pixel selection and replacement is applied at every epoch, changing both the location and values of the selected active pixels. Unlike \cite{Krull2019} who randomly select the replacement pixel value from within the full neighbourhood region, we explicitly ensure that an active pixel's value cannot be replaced by itself.

This `corrupted' version of the raw noisy data becomes the input to the network. For the conventional, self-supervised case, the network's target are the raw noisy data themselves. The purpose of the network is to learn to predict an active pixel's value based on the contribution of neighbouring pixels. Therefore, the loss is only calculated at the active pixels' locations,
\begin{equation}
    \mathcal{L}_{denoising} = \frac{1}{N_s N_p} \sum_{j=1}^{N_s} \sum_{i=1}^{N_p} |x^j_i - \hat{y}^j_{i} | ,
\end{equation}
where $N_s$ represents the number of training samples, $N_p$ denotes the number of blind-spots per sample, $x^j_{i}$ is the original value of the $i$-th active pixel, and $\hat{y}^j_{i}$ is the predicted value of the $i$-th active pixel. Here, the operator $|\cdot|$ denotes the absolute value. As with most NN tasks, the loss is back-propagated and used to update the weights \cite{Siddique2001}. Under the assumption of i.i.d. noise, the network cannot learn to accurately recreate the noise component of each active pixel due to its random nature. Therefore, the loss function is minimised across the training samples as the signal component is being recreated without predicting the noise.

The blind-spot procedure has a number of parameters, in particular the number of active pixels and the neighbourhood radius from which the replacement pixel value is selected. Previous work by \cite{Birnie2021} highlighted how increasing both those values can allow some accommodation for breaking the i.i.d. noise assumption. Table \ref{tab:N2Vparams} compares the different parameters selected for different studies. In this work, we aim to tackle the full noise field, a combination of random and coherent noises, therefore we initialised our parameter selection based on the seismic field data parameters of \cite{Birnie2021}. Manual tuning of the parameters did not provide any improvement in the networks' denoising performance.

\begin{table}[!htb]
\centering
\caption{Blind-spot network parameters for different data and noise types. }
\label{tab:N2Vparams}
\begin{tabular}{|c|c|c|c|c|}
\hline
\textbf{Study} & \textbf{Data Type} & \textbf{Noise Type} & \textbf{\% Active Pixels} & \textbf{Neighborhood Radius} \\ \hline
Krull et al., 2019 & Natural Images & i.i.d. & 0.2 - 2 & 5 \\ \hline
Birnie et al., 2021 & Seismic & WGN & 0.2 & 5 \\ \hline
Birnie et al., 2021 & Seismic & Bandpassed & 25 & 15 \\ \hline
Birnie et al., 2021 & Seismic & Field Data & 33 & 15 \\ \hline
This study & Seismic & Field Data & 33 & 15 \\ \hline
\end{tabular}
\end{table}

\subsection{Network design and training procedures}
Throughout the experiments shown in this paper, the network design and training parameters are held constant. These are detailed in Table \ref{tab:NNparams} and were chosen based on previous blind-spot network implementations combined with a trial-and-error optimisation for the field dataset considered in this study.

\begin{table}[!htb]
\centering
\caption{The neural network architecture and training parameters.}
\label{tab:NNparams}
\begin{tabular}{|c|c|c|c|}
\hline
\textbf{NN Architecture} & \textbf{\# Layers} & \textbf{\# Initial Filters} & \textbf{Kernel Size} \\ \hline
UNet & 2 & 32 & 3x3 \\ \hline
\textbf{Loss Function} & \textbf{Batch Size} & \textbf{Optimiser} & \textbf{Initial LR} \\ \hline
MAE & 32 & Adam & 1e-4 \\ \hline
\end{tabular}
\end{table}

For the self-supervised, blind-spot implementations, the number of epochs over which the network is trained has been shown to be a determining factor in the networks ability to suppress the noise. The longer the network is trained, the more time it has to learn to replicate the noise, as well as the signal. The selection of the number of epochs to train over is strongly dependent on the signal and noise within the data of interest. As such, in this study we investigate the optimum number of epochs for the self-supervised networks.

\subsection{Transfer learning implementation}
This study lies its foundations on the hypothesis that a network can be initially trained to learn to predict the signal component of an active pixel without the network learning to also replicate the noise component. This initial training is performed using overly-simplistic synthetic datasets with noisy-clean training pairs prior to the learnings being transferred to the self-supervised procedure via a weight transfer. Throughout we will refer to the supervised (initial) training as the base-training and the subsequent self-supervised training, as the fine-tuning.

Figure \ref{fig:workflow} represents the proposed workflow for the inclusion of the base-training (left-side) prior to the original self-supervised training (right-side). Both networks are trained in a blind-spot manner, i.e., following the active pixel selection and corruption of \cite{Krull2019}. Unlike for self-supervised, blind-spot procedures, the target of the base-training are the clean, synthetic data. This clean target provides the network with the exact values that we wish for it to predict, i.e., only the signal component, but teaches it to do so from only the neighboring pixels 

For the fine-tuning, this is performed as a self-supervised task under the assumption that no clean target data are available. As such, only a blind-spot implementation is utilised and the target is the raw, noisy data themselves. The initial weights of the networks are transferred from the earlier base-training experiments prior to being optimised following the self-supervised training procedure, as illustrated by the long, black arrow in Figure \ref{fig:workflow}.

For comparative purposes, a self-supervised, blind-spot network is trained from randomised weights, as opposed to using the transferred weights from the base-training. In addition, the blind-spot network trained in a supervised manner is also applied to the data to provide a benchmark against supervised options.

%%%%%%%%%%%%%%%%%%%%%%%%%%%%%
\section{Training data generation}
\begin{figure}[ht]
  \centering
  \includegraphics[width=0.75\textwidth]{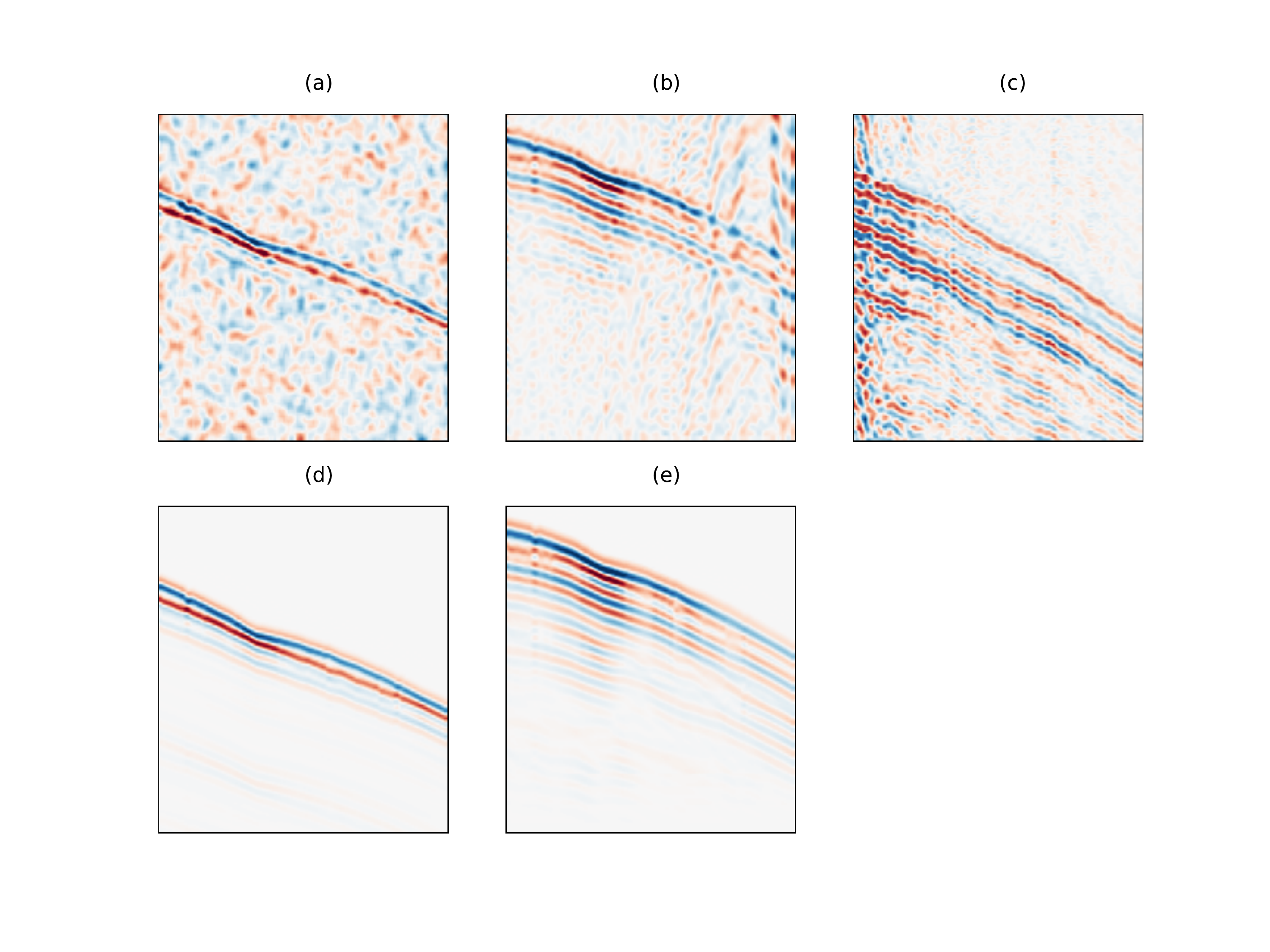}
  \caption{Microseismic events from synthetic (a,d), semi-synthetic (b,e) and field data (c). Top row illustrates the data to be denoised, whilst the bottom row illustrates the noise-free target data (i.e., the generated waveforms).}
  \label{fig:data}
\end{figure}

Self-supervised learning procedures aim to tackle the often unobtainable requirement of having clean-noisy training pairs. In geophysics this requirement has previously been overcome, to a certain extent, by the use of realistic synthetic datasets, which are often non-trivial to generate - requiring expensive waveform modelling and realistic noise generation. To avoid re-introducing this non-triviality into the blind-spot procedure we use simplistic synthetic datasets for the supervised training. In this study we focus on a passive seismic dataset previously analysed by \cite{Wang2021}. Three different datasets utilised in this study as illustrated in Figure \ref{fig:data}.

The synthetic data are generated using a two-step procedure: first,  Eikonal-based modelling is used to compute the travel-times at all receivers to a given subsurface volume and then, the synthetic seismograms are generated via convolutional modelling. Receivers are placed on the surface and distributed in a similar manner to those used to acquire the field data. Travel-times are computed from each receiver to all subsurface points through a laterally homogeneous velocity model using the fast marching method \cite{Sethian1999}, implemented via the scikit-fmm python package \cite{Furtney2019}. For memory purposes, only the travel-times corresponding to a specific area of interest (i.e., around the reservoir) are retained for generating the synthetic seismograms. After the computation of the travel-time tables, source-locations are randomly selected and the corresponding synthetic seismic data is generated per receiver by placing an Ormsby wavelet at the arrival time. In order to increase variety within the training data, the frequency content of the wavelet is randomly chosen per event, within a reasonable frequency range for microseismic events. Finally, to account for geometric spreading, each trace is scaled by $1/r$, where $r$ represents the distance between the source and corresponding receiver.

Using an Eikonal solver can drastically decrease the compute cost, whilst the reliance on prior geological knowledge is also reduced through the use of a laterally, homogeneous model. Through this approach, we do not account for realistic source mechanisms or complex propagation. Such assumptions contradict our knowledge of elastic waveform propagation from microseismic sources, rendering the generated waveforms highly unrealistic. In order to introduce noise, Coloured, Gaussian Noise (CGN) is added to all the waveforms completing the synthetic training datasets generation. Whilst CGN is slightly more realistic than the commonly used White, Gaussian Noise (WGN), it still exhibits highly unrealistic noise properties in comparison to noise observed in field data.

To allow for a thorough investigation of the benefits of the proposed methodology, a realistic semi-synthetic dataset is generated. The waveforms are excited by point sources randomly placed in the same reservoir region as before, however this time the waveforms propagate through a 3D velocity model (from \cite{Wang2021}) using an elastic wave equation, implemented in Madagascar \cite{Fomel2003}. To include realistic noise, noise from the field data is directly added to the modelled, elastic waveform data. 

To increase data volumes, giving us an adequate number of training samples, for both the synthetic and semi-synthetic datasets, waveforms are randomly flipped (i.e., polarity reversed) with arrival times shifted across the recording window. For the field dataset, all available events are used for training the network prior to its application onto the same field events. As this is a self-supervised procedure with no clean training target available, we do not have the same over-fitting concerns as supervised procedures and therefore, there is no requirement for a blind, holdout set.

\section{Results}

\subsection{Semi-synthetic examples}
As there is no ground-truth available when denoising field data, an initial denoising experiment is performed where semi-synthetics represent the field data. This allows us to perform a rigorous statistical analysis on the denoising products identifying the volume of noise suppressed/remaining, as well as any signal damage encountered. The Peak Signal-to-Noise Ratio (PSNR) is used throughout as a measurement of the overall denoising performance. Signal leakage is a huge concern in seismic-, and particularly microseismic-, processing with geophysicists typically preferring to leave in more noise than to cause any damage to the signal \cite{Mousavi2016a}. To quantify signal leakage, the Mean, Absolute Error (MAE) is computed between the clean and denoised data on the pixels where the clean waveform is present. Finally, to compute the approximate volume of noise suppressed, the MAE is computed between the additive noise (i.e., subtracting the clean data from the noisy data) and denoised data on the pixels everywhere that a clean waveform is not present.

\begin{figure}[ht]
  \centering
  \includegraphics[width=0.75\textwidth]{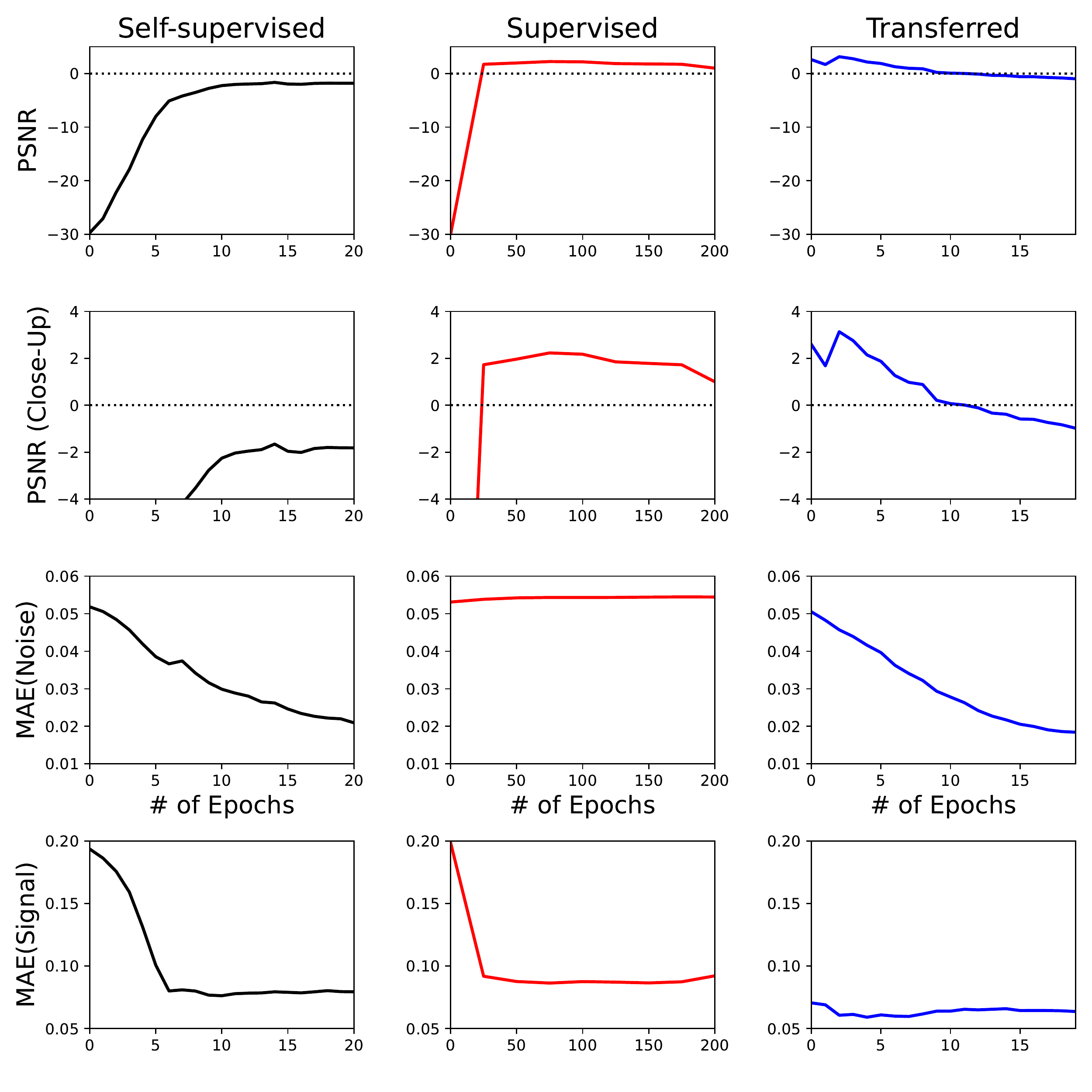}
  \caption{Denoising performance with respect to epochs for self-supervised (left), supervised (center) and fine-tuned (right) networks. The top two rows illustrate the PSNR change through denoising with different y-axes scales for comparison, the third illustrates the noise suppressed whilst the bottom row illustrates the volume of signal leakage occurring.}
  \label{fig:epochprogression}
\end{figure}

As previously mentioned, three blind-spot networks are trained: one self-supervised trained on the `field' data, one trained in a supervised manner on the synthetic data, and one fine-tuned in a self-supervised manner, initialised with the weights of the supervised model. To identify the optimum number of training epochs per model, a statistical analysis is performed at regular checkpoints to investigate the denoising performance. Note, self-supervised methods require substantially less epochs as they will quickly learn to replicate coherent noise if trained for too long. Figure \ref{fig:epochprogression} illustrates the different networks' progression during their respective training cycles. An interesting observation is that the volume of noise suppressed with the transferred approach, closely follows the trend of the self-supervised denoising procedure: decreasing the volume of noise suppressed per epoch as the network starts learning to replicate the noise as well as the signal. However, the amount of signal leakage is substantially less than the other two training approaches, even in the early epochs - highlighting that the fine-tuning stage has quickly learnt to adapt to the field source signature.

Considering the trends in each networks' performance, and verified via a visual analysis of the denoised products we use the following number of epochs for the different models: 
\begin{itemize}
    \item \textbf{Self-Supervised}: 15 epochs,
    \item \textbf{Supervised}: 200 epochs, and
    \item \textbf{Transferred}: 200 supervised plus 2 self-supervised epochs.
\end{itemize}

\begin{figure}[ht]
  \centering
  \includegraphics[width=1.\textwidth]{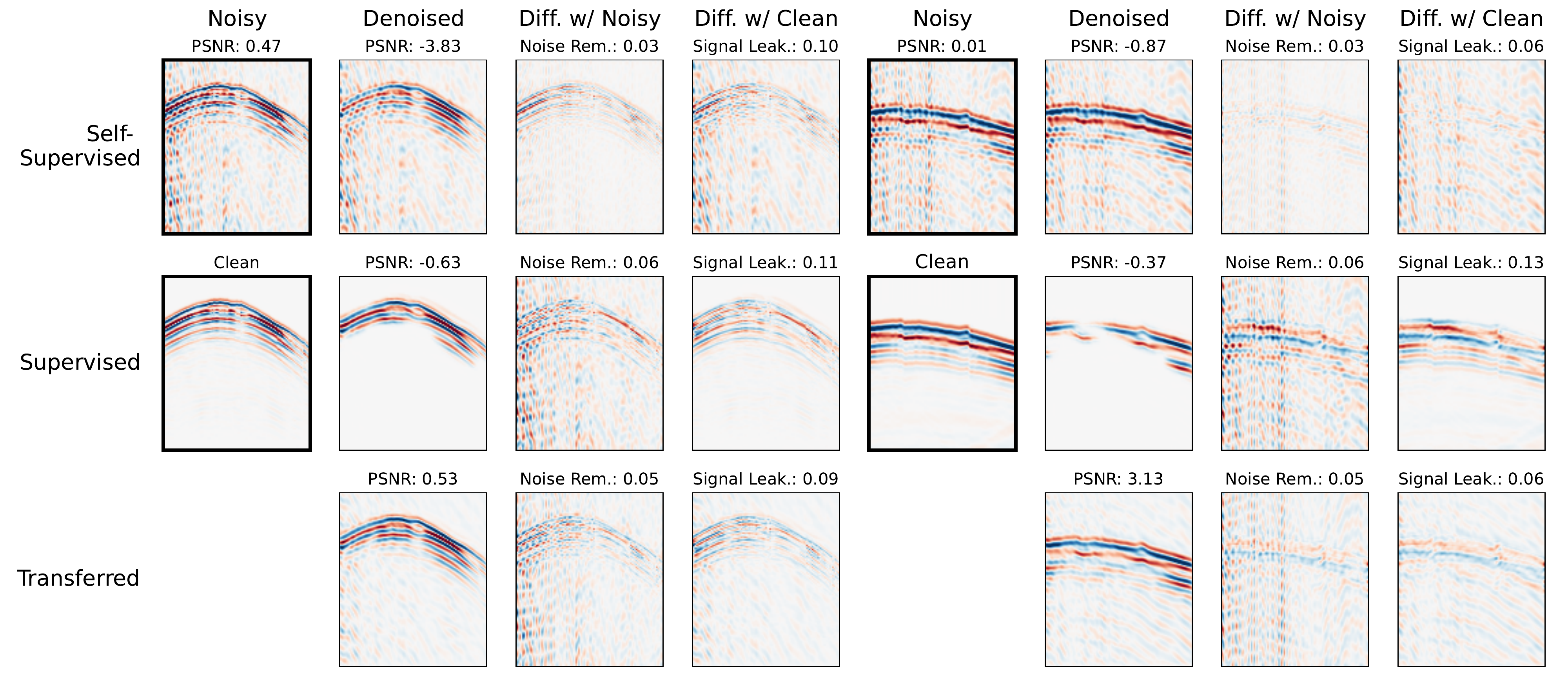}
  \caption{Two semi-synthetic events denoised by the three different networks: self-supervised (top row), supervised (middle row) and the transfer learning network (bottom row). The first and fifth columns illustrate the noisy and clean data, while the second and sixth display the denoised products. The next two columns show the difference between the noisy and clean data, respectively, for the denoising products of the different networks. Values for the PSNR, noise removed and signal leakage are given in each subplots titles.}
  \label{fig:semisynth_images}
\end{figure}

\begin{figure}[ht]
  \centering
  \includegraphics[width=0.75\textwidth]{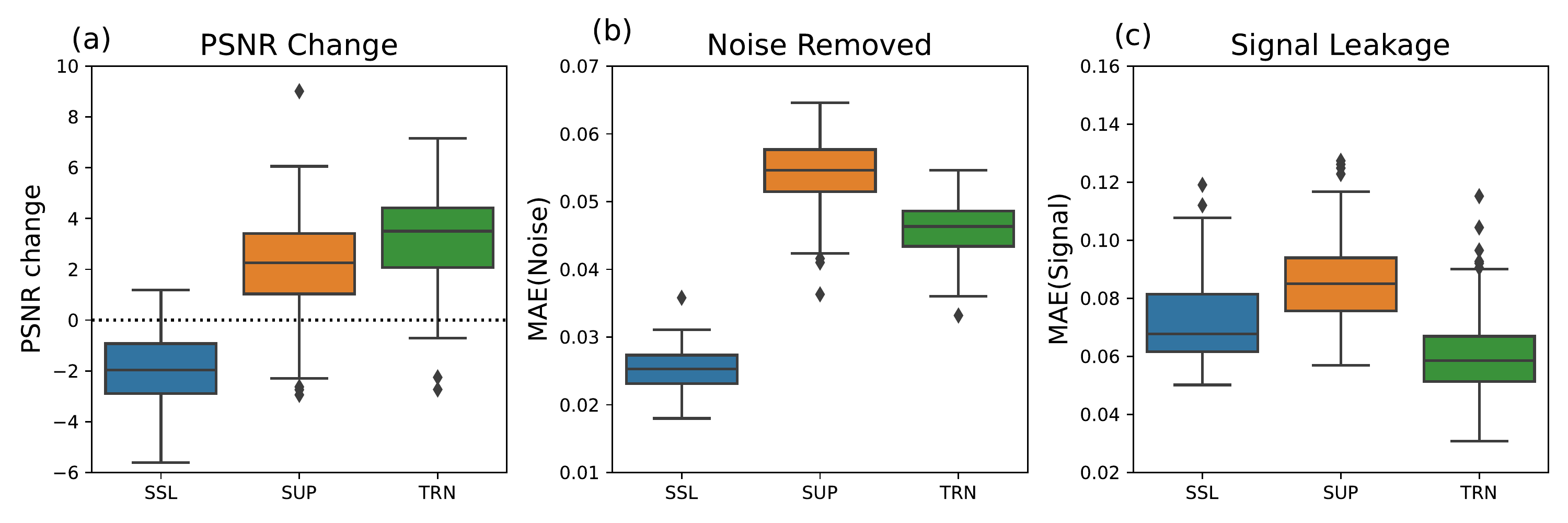}
  \caption{Comparison of denoising performance of the different networks: fully self-supervised (SSL), fully-supervised (SUP), and the network pre-trained in a supervised manner prior to self-supervised, fine-tuning (TRN). The box plots show for 100 events the (a) overall change in PSNR, (b) volume of noise removed, and (c) shows the amount of signal leakage encountered. The dashed line in (a) indicates a benchmark against the PSNR of the original, noisy data.}
  \label{fig:semisynth_stats}
\end{figure}

The trained models were tested on 100 new events, unseen in the training and hyper-parameter tuning procedures. Figure \ref{fig:semisynth_images} portrays the performance of the three different models on two of these microseismic events, with noticeably different frequency content and moveout shapes. In both events, the proposed methodology incorporating transfer learning results in the highest PSNR and the least amount of signal leakage. The network trained in a supervised manner suppresses all noise within the data, something neither of the techniques including self-supervised learning achieve. However, the signal damage (/leakage) is significant, particularly in the second event. Furthermore, Figure \ref{fig:semisynth_stats} illustrates the change in the PSNR for the 100 investigated events (a), alongside the volume of noise suppressed (b) and the signal leakage occurring (c). Similar to the results from the two displayed events (in Figure \ref{fig:semisynth_images}), the transfer learning approach offers the highest PSNR improvement and the least amount of signal damage. Whilst the transferred approach cannot match the noise removal of the supervised procedure, substantial noise reduction is observed in comparison to the standard, self-supervised approach. 

\subsection{Field examples}
\begin{figure}[ht]
  \includegraphics[width=1.\textwidth]{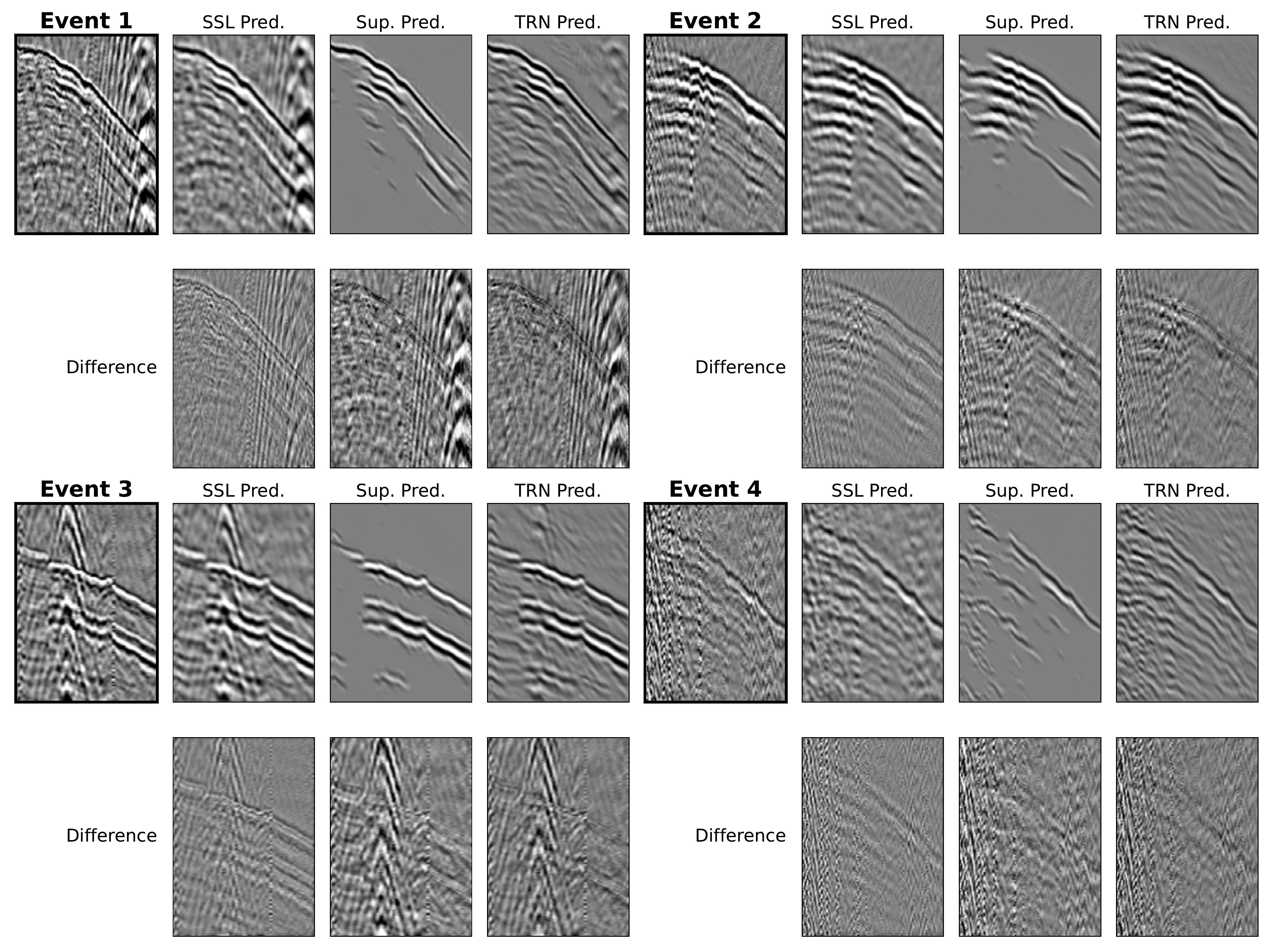}
  \caption{Comparison of the different denoising networks' performance on four different microseismic events observed within field data. The top, left panel of each event shows the raw noisy data, with the remaining panels in the top rows showing the denoising results for the different networks. The bottom rows show the difference between the denoised products and the raw, noisy events.}
  \label{fig:field_results}
\end{figure}

The same supervised model is applied to the field data, whilst a new self-supervised training is implemented on a similar model initiated randomly and initiated with the supervised model weights for what we previously identified as the optimum number of epochs. Figure \ref{fig:field_results} illustrates the performance of the three different networks on four microseismic events observed in the field dataset. As in the semi-synthetic examples, the events have different frequency content, move-out patterns and noise properties. As there is no noise-free equivalent of the field events, quantifying the increase in the SNR and the amount of signal damage is non-trivial. However, qualitatively we can observe from the top row of plots for each event that the transfer learning procedure suppressed substantially more noise than the self-supervised procedure. In particular, the high-energy noise observed on a number of traces in both Event One and Three has been greatly reduced. In addition to this, by considering the difference between the noisy data and denoised products (bottom rows), it is clear that the transfer learning procedure introduces the least amount of signal leakage.

\begin{figure}[ht]
  \includegraphics[width=1.\textwidth]{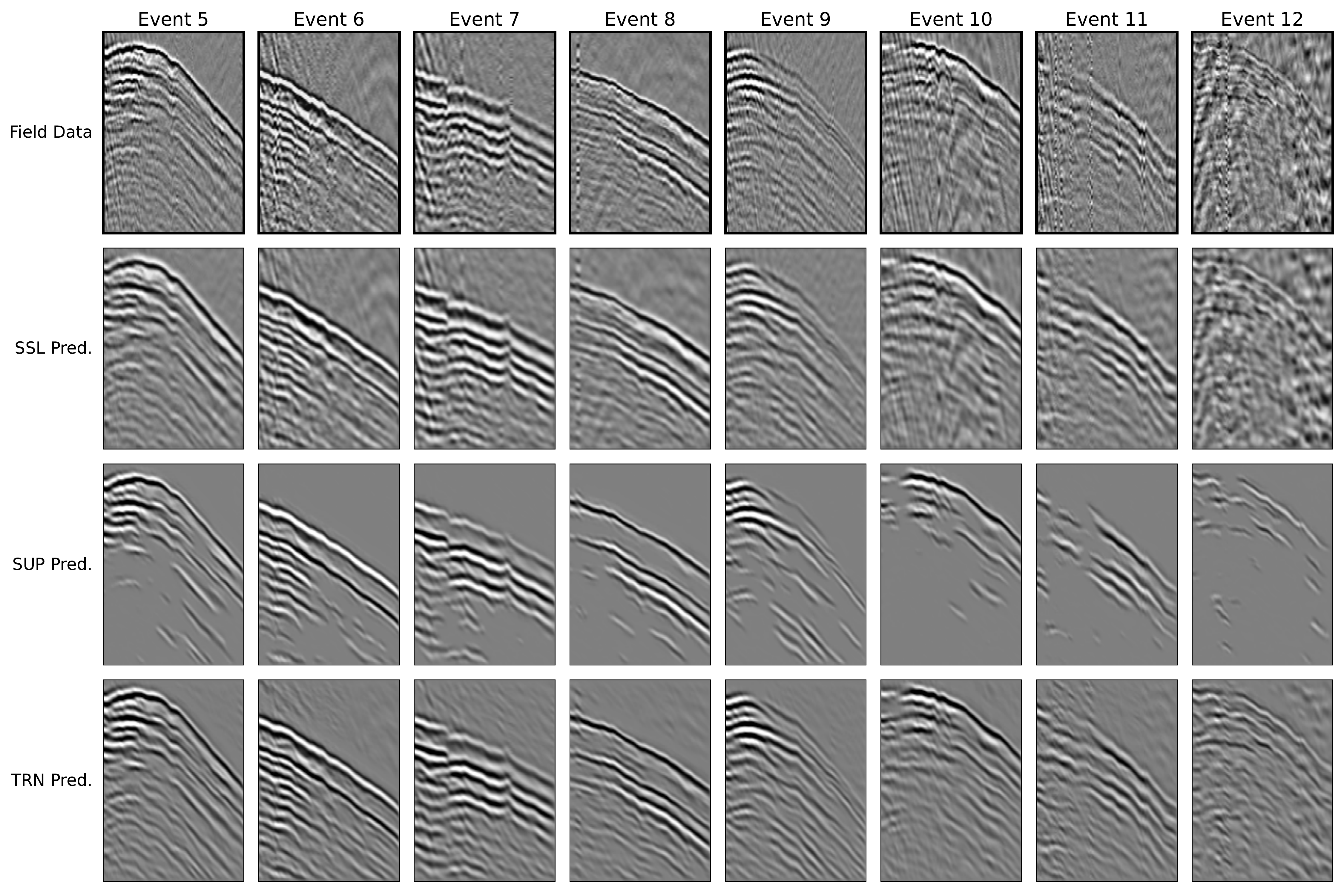}
  \caption{Additional comparison of denoising performances on field data. The top row portrays the raw field data whilst the remaining three rows represent the denoised products from the self-supervised network, supervised network, and transferred network, respectively.}
  \label{fig:field_results2}
\end{figure}

The denoising performance for the different approaches is further illustrated in Figure \ref{fig:field_results2} for an additional eight events, with varying SNRs and arrival move-outs. The first four events have relatively large SNRs, with all three denoising methods managing to remove some noise whilst retaining a relatively clean signal component. Although, for the fourth event the supervised method is beginning to noticeably struggle with the signal reconstruction. The final four events are contaminated by higher noise levels. In this instance, the supervised method cannot accurately reproduce the signal whilst the fully self-supervised procedure leaks significant noise into the denoised product. The fine-tuned model (i.e., transferred approach) is shown to provide a balance between the drastic signal damage of the supervised procedure and the noise inclusion of the self-supervised approach.

\section{Discussion}
Blind-spot, self-supervised denoising techniques, such as N2V \cite{Krull2019} and StructN2V \cite{Broaddus2020}, remove the common requirement of clean-noisy pairs of data for training a denoising neural network. However, the main drawback of N2V is the random noise assumption, that rarely holds for noise in seismic data. \cite{Birnie2021} illustrated how training on data with even minor correlation within the noise field results in the network learning to reproduce both the desired signal and noise. Whilst N2V's successor, StructN2V, can suppress coherent noise, it requires a consistent noise pattern for which a specific noise mask is built, e.g., masking the noise along a specific direction. This has shown great promise for trace-wise noise suppression, such as dead sensors \cite{Liu2022}  or blended data \cite{Luiken2022}, however it is not practical for the suppression of the general seismic noise field which is continuously evolving. In this work, we have proposed to initially train a network on simplistic synthetic datasets and then fine-tune the model in a self-supervised manner on the noisy field data. This base-trained model has learnt to replicate only the signal component of a central pixel from noisy neighbouring pixels. Therefore, a very small number of epochs (e.g., two) is needed for the network to adapt to the field seismic signature, significantly reducing the amount of time the network is exposed to the field noise. As such, the final network is capable of predicting the field seismic signature without substantial recreation of the field noise.

\subsection{Frugal synthetic dataset generation}
The synthetic datasets generated for training only assumed a rough vertical profile of the subsurface is known. As such, there was no requirement of a detailed subsurface model or of the expected noise properties. In addition to requiring less prior knowledge, generating the travel-timetables for the subsurface region of interest took three minutes per receiver with scikit's Fast Marching Method (scikit-fmm - \cite{Furtney2019}) using a single core [Intel(R) Xeon(R) W-2245 CPU @ 3.90GHz]. Subsequently, synthetic seismograms were generated from the travel-timetables using convolutional modelling. Allowing for the random selection of the wavelet's frequency content, the arrival time shift and the generation of the wavelet, the convolutional modelling step took $1.5$ seconds per microseismic event. In comparison, the 3D wavefield modelling for the semi-synthetic dataset took $1.75$ hours per shot, due to requirements on the time and space sampling to ensure stability. Whilst these computation times may not represent the performance of state-of-the-art computational modelling software, the difference in compute time between 2D versus 3D and acoustic versus elastic modelling is well known.

Another benefit of this approach is the removal of the requirement for the inclusion of realistic noise. A number of studies in deep learning have circumvented this by the inclusion of previously recorded noise (e.g., \cite{Mousavi2019, Zhu2020, Wang2021}). However, this requires a substantial amount of pre-recorded data and is restricted to the geometry/field where the data were collected. An alternative is to consider statistical noise modelling procedures (e.g., \cite{Birnie2016}) or perform waveform modelling (e.g., \cite{Dean2015}). However, both of these methods can be computationally expensive, and do not guarantee a perfect recreation of the noise properties within an unseen field dataset. Therefore, neither provide an ideal alternative for incorporating realistic noise into the training dataset.

One final benefit we see is that both the synthetic data and the pre-trained, base-model are not closely coupled to any specific well-site condition. For example, if using a fully supervised model trained with a noise model assuming no well-site activity, a new training dataset and model would need to be created for use on data collected during periods of injection. By utilising an initial, unrealistic noise model, the synthetic dataset for the base-training is not highly correlated to the true field conditions. Therefore, allowing the field conditions to be transferred into the denoising procedure at the fine-tuning stage - fully driven by the field data. As such, the only re-training required due to a change in site conditions would be the two-epoch fine-tuning stage - a matter of mere seconds.

\subsection{Use of blind-spots in base-training}
\begin{figure}[ht]
  \centering
  \includegraphics[width=0.75\textwidth]{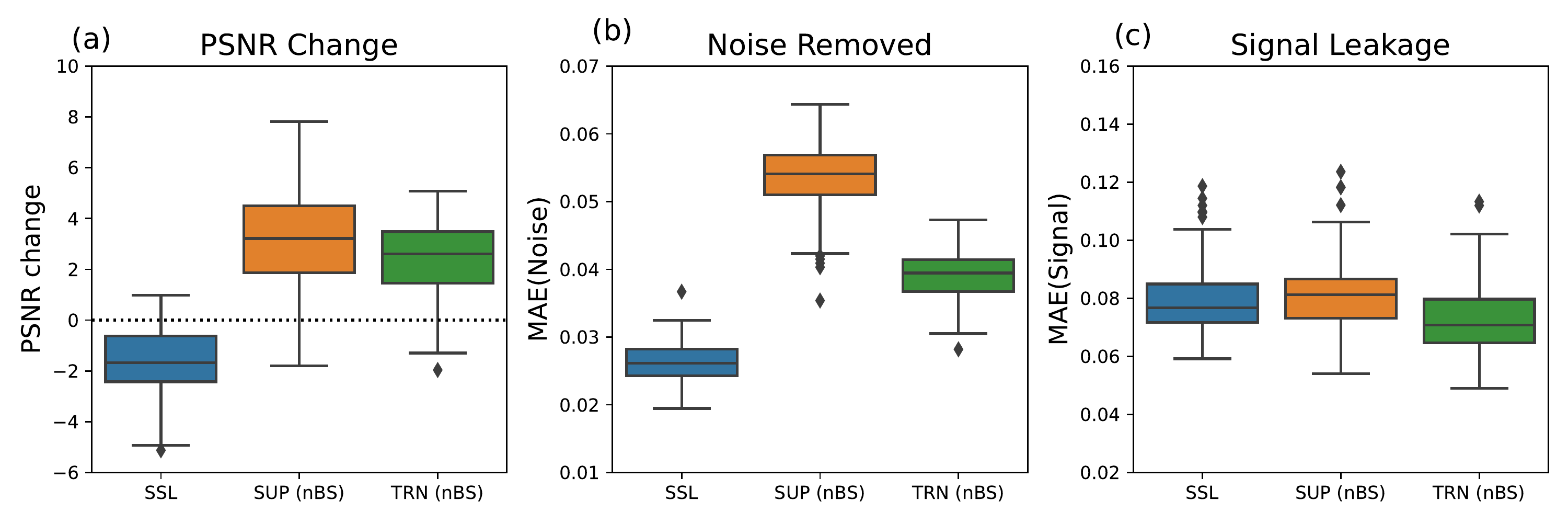}
  \caption{Comparison of denoising performance when the supervised network is trained in a conventional manner, i.e., when the blind-spot is only implemented for the self-supervised training stages. The box plots show for 100 events the (a) overall change in PSNR, (b) volume of noise removed, and (c) shows the amount of signal leakage encountered. The dashed line in (a) indicates a benchmark against the PSNR of the original, noisy data.}
  \label{fig:noBS_stats}
\end{figure}

The hypothesis of this project was that initially training the network to learn the signal component of a pixel based off neighbouring pixels' values would allow faster training times at the self-supervised stage and therefore, allowing the network less time to learn to predict the coherent components of the field data's noise. As such, it is important to implement the pre-training as a blind-spot scheme through the standard identification and corruption of active pixels, and the loss computation on only the active pixels location. Following a conventional supervised training scheme, going from noisy to clean without blind-spots, would teach the pre-trained network to utilise the central pixel's value when computing itself - something not possible in the self-supervised, blind-spot fine-tuning. To quantify the importance of consistent use of blind-spots throughout the training procedures, Figure \ref{fig:noBS_stats} highlights the difference in the performance between a network pre-trained with blind-spots versus one pre-trained in the conventional deep learning manner. In comparison to Figure \ref{fig:semisynth_stats}, it is clear that the performance of the new fine-tuned network no longer outperforms the supervised network. However, the original, fine-tuned network from earlier in the study still outperforms the supervised network. Note, in this instance the supervised network has not been implemented in a blind-spot manner. This result is unsurprising as when trained in a conventional manner pixels can be used to predict themselves. Therefore, during the self-supervised, fine-tuning stage not only did the network learn to recreate the field signal but it also had to learn that it cannot use the pixel in its prediction.

\section{Conclusion}
Self-supervised procedures for seismic denoising have typically targetted a single component of the noise field, either random noise or noise coherent along a specific axis. In this study, we proposed a methodology to allow the suppression of the full noise field by the inclusion of supervised base-training. This initial training is performed using highly simplistic synthetic datasets generated from an acoustic waveform modelling through a laterally homogeneous subsurface model. Benchmarked on a semi-synthetic dataset, the proposed network is shown to increase the volume of noise suppressed in comparison to a standard self-supervised network, whilst the quantity of signal damage is substantially less in comparison to a network trained in a supervised manner. Similar conclusions are drawn when the networks are adapted for microseismic events observed in field data. Our proposed procedure of supervised base-training on a simplistic synthetic dataset prior to self-supervised fine-tuning is repeatedly shown to provide the best denoising performance.

\section{Acknowledgements}
The authors thank the KAUST Seismic Wave Analysis Group for insightful discussions. For computer time, this research used the resources of the Supercomputing Laboratory at King Abdullah University of Science \& Technology (KAUST) in Thuwal, Saudi Arabia. 

%% References and Citations:
% \section{References}
\typeout{}
\bibliographystyle{unsrt} 
\bibliography{bibliography}

\end{document}